\documentclass[prl,twocolumn,superscriptaddress,showpacs]{revtex4}

\usepackage{graphicx}
\usepackage{dcolumn}
\usepackage{bm}

\begin{document}

\preprint{ATB-1}

\title{Spin correlations among the charge carriers in an ordered stripe
phase}

\author{A.T. Boothroyd}
\email{a.boothroyd1@physics.ox.ac.uk}
\homepage{http://xray.physics.ox.ac.uk/Boothroyd}\affiliation{
Department of Physics, Oxford University, Oxford, OX1 3PU, United
Kingdom }
\author{P.G. Freeman}\affiliation{
Department of Physics, Oxford University, Oxford, OX1 3PU, United
Kingdom }
\author{D. Prabhakaran}\affiliation{
Department of Physics, Oxford University, Oxford, OX1 3PU, United
Kingdom }
\author{A. Hiess}\affiliation{
Institut Laue-Langevin, BP 156, 38042 Grenoble Cedex 9, France }
\author{M. Enderle}\affiliation{
Institut Laue-Langevin, BP 156, 38042 Grenoble Cedex 9, France }
\author{J. Kulda}\affiliation{
Institut Laue-Langevin, BP 156, 38042 Grenoble Cedex 9, France }
\author{F. Altorfer}\affiliation{
Laboratory for Neutron Scattering, ETH Zurich and PSI Villigen,
CH-5232 Villigen PSI, Switzerland }
\date{\today}

\begin{abstract}
We have observed a diffuse component to the low-energy magnetic
excitation spectrum of stripe-ordered La$_{5/3}$Sr$_{1/3}$NiO$_4$
probed by neutron inelastic scattering. The diffuse scattering
forms a square pattern with sides parallel and perpendicular to
the stripe directions. The signal is dispersive, with a maximum
energy of $\sim 10$\,meV. Probed at 2\,meV the scattering
decreases in strength with increasing temperature, and is barely
visible at 100\,K. We argue that the signal originates from
dynamic, quasi- one-dimensional, antiferromagnetic correlations
among the stripe electrons.
\end{abstract}

\maketitle


The occurence of stripe correlations in superconducting cuprates
and other doped antiferromagnets is well established from
extensive experimental studies \cite{Stripes-expt} following
earlier theoretical work \cite{Stripes-theory}. The stripe phase
is characterized by a segregation of charge carriers into narrow
channels that act as antiphase domain walls separating
antiferromagnetic regions of the host spins. Several theoretical
scenarios have been presented in which a spin gap and pairing
instabilities appear among the stripe electrons, opening the door
to superconductivity \cite{Stripes-SC-theory}. These findings
suggest that stripes could play an important role in the physics
of high-$T_{\rm c}$ superconductivity.

Given the characteristic spin and charge correlations associated
with local stripe order it is natural to search the spin
excitation spectrum for clues to the relationship between stripes
and superconductivity. Neutron inelastic scattering has revealed a
number of interesting features in the spin dynamics of the cuprate
superconductors, some of which, e.g. excitations at incommensurate
wavevectors \cite{cuprates-stripes}, have been interpreted as
evidence for dynamic stripe correlations
\cite{Tranquada-Nature-1995}. At the same time, the characteristic
spin excitation spectrum of ideal stripes has been probed in
studies of non-superconducting compounds with well defined stripe
order, especially La$_{2-x}$Sr$_x$NiO$_{4+\delta}$
\cite{INS-stripes-nickelates,Boothroyd-PRB-2003}. In none of these
previous studies, however, has any evidence been found for spin
correlations among the stripe electrons themselves.


In this paper we report neutron inelastic scattering measurements
on La$_{5/3}$Sr$_{1/3}$NiO$_4$ which reveal a pattern of inelastic
diffuse scattering consistent with the existence of quasi-
one-dimensional (1D) antiferromagnetic (AF) correlations parallel
to the stripe direction. The correlations are found to be dynamic,
with a maximum energy $\sim 10$\,meV. The results provide the
first evidence for spin correlations among the charge carriers in
a stripe phase.

The experiments were performed on single crystals of
La$_{5/3}$Sr$_{1/3}$NiO$_4$ grown by the floating-zone method
\cite{Prabhakaran-JCG-2002}. Neutron inelastic scattering data
were collected on the triple-axis spectrometers IN8, IN20 and IN22
at the Institut Laue-Langevin, and RITA-II at the Paul Scherrer
Institut.  Most of the data reported here were obtained with a
fixed scattered neutron energy $E_{\rm f} = 14.7$\,meV and with a
graphite filter installed after the sample to suppress higher
orders. The exception is the inelastic scan shown in
Fig.\,\ref{fig:2}(a), which was measured with an incident energy
$E_{\rm i} = 14.7$\,meV and with the filter before the sample. The
incident and final neutron energies were selected by Bragg
reflection from the $(002)$ planes of graphite. When neutron
polarization analysis was required Heusler $(111)$ was used for
the monochromator and analyzer.

Stripe ordering in La$_{5/3}$Sr$_{1/3}$NiO$_4$ is observed below
$T_{\rm SO}\simeq 200$\,K, and consists of parallel lines of doped
holes separating AF-ordered bands of Ni$^{2+}$ spins ($S=1$)
\cite{LaSrNiO-x_1/3}. The stripes run diagonally across the square
lattice formed by the Ni sites on the NiO$_2$ layers. As there is
no reason to favour one diagonal over the other a bulk sample
contains an equal proportion of spatially-separated domains with
stripes along each diagonal.


Previous measurements have shown that the spin excitation spectrum
of La$_{5/3}$Sr$_{1/3}$NiO$_4$ is dominated by sharp propagating
modes of the host Ni$^{2+}$ spins \cite{Boothroyd-PRB-2003}. These
spin-wave-like excitations are highly 2D in nature, and extend in
energy up to $\sim 80$\,meV. They disperse from the stripe
superlattice zone centres, which have 2D wavevectors ${\bf Q}_{\rm
s} = (h+\frac{1}{2},k+\frac{1}{2}) \pm (\frac{1}{6},\frac{1}{6})$
for one stripe domain and $(h+\frac{1}{2},k+\frac{1}{2}) \pm
(-\frac{1}{6},\frac{1}{6})$ for the other, where $h$ and $k$ are
integers indexed with respect to the body-centred tetragonal
lattice (cell parameters $a=3.8\,{\rm \AA}$ and $c=12.7\,{\rm
\AA}$).

\begin{figure}
\includegraphics{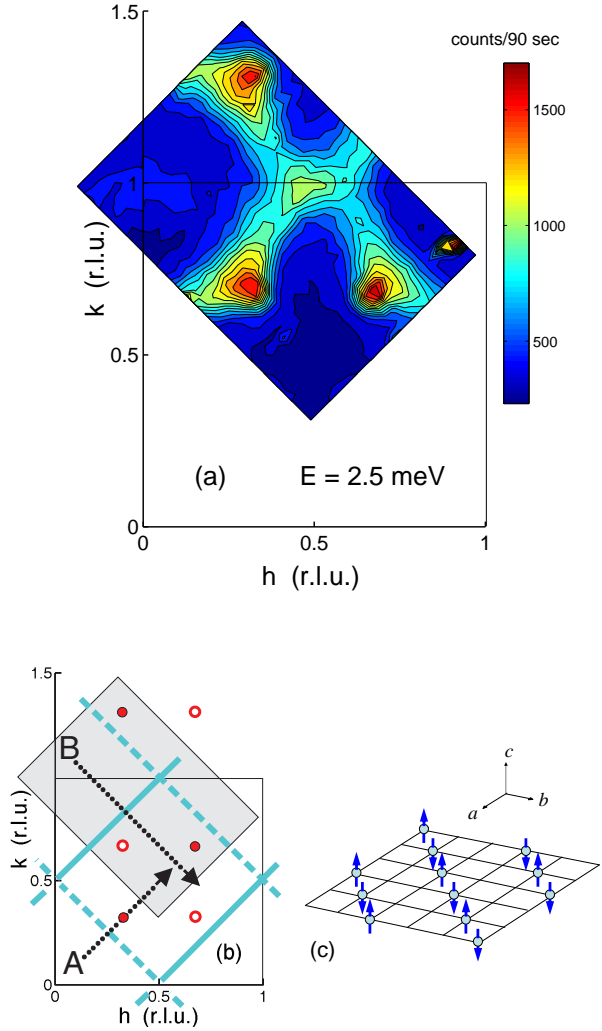}
\caption{(Color online). Diffuse neutron inelastic scattering from
La$_{5/3}$Sr$_{1/3}$NiO$_4$. (a) Contour plot of the intensity in
the $(h,k,0)$ plane in reciprocal space at an energy transfer of
2.5\,meV. The plot is derived from a $25\times 35$ rectangular
grid of data points. (b) Diagram of the $(h,k,0)$ plane in
reciprocal space. Circles are stripe superlattice zone centres.
Diagonal lines indicate where diffuse scattering would be observed
if the stripes behaved as 1D AF chains. Filled circles and full
lines are for $[1, -1, 0]$ stripe domains; open circles and broken
lines are for $[1, 1, 0]$ stripe domains. The dotted lines marked
A and B are scan directions used in the experiment. (c) $[1, 1,
0]$ stripe domain showing the square lattice of Ni sites with
local 1D AF correlations on the stripes, as proposed here.
\label{fig:1} }
\end{figure}

\begin{figure}
\includegraphics{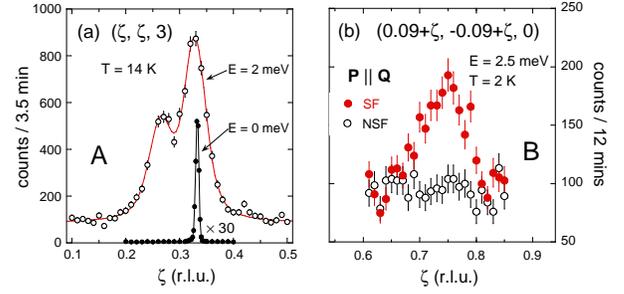}
\caption{(Color online). Constant-energy scans along lines
equivalent to those marked A and B in Fig.\,\ref{fig:1}(b). The
polarized-neutron data in (b) show the spin-flip (SF) and
non-spin-flip (NSF) channels recorded with the neutron
polarization parallel to the scattering vector. \label{fig:2}}
\end{figure}

In the present work we explored the low energy part of the
spectrum over a more extended region of reciprocal space.
Fig.\,\ref{fig:1}(a) shows the variation in scattered intensity
over a rectangular portion of the $(h,k,0)$ plane in reciprocal
space, measured at an energy of 2.5\,meV and at a temperature of
2\,K. The region of reciprocal space covered by the measurements
is shown in Fig.\,\ref{fig:1}(b). The intensity map contains
localized `hot spots' of intensity centred on ${\bf Q}_{\rm s}$
points such as $(\frac{1}{3},\frac{2}{3}),
(\frac{2}{3},\frac{2}{3})$, etc, marked as circles on
Fig.\,\ref{fig:1}(b). This scattering is from spin wave
excitations of the host Ni$^{2+}$ spins, as described above.

More interestingly, the data in Fig.\,\ref{fig:1}(a) also show a
pattern of diffuse scattering distributed around the diagonal
lines drawn on Fig.\,\ref{fig:1}(b). The diffuse scattering does
not pass through the stripe zone centres, and is in fact separated
from them. This is illustrated in Fig.\,\ref{fig:2}(a), which
contains two scans measured along a line equivalent to that marked
A in Fig.\,\ref{fig:1}(b). In the 2\,meV scan there are two
resolved peaks, one centred on $(0.27,0.27)$, from the diffuse
scattering, and the other centred on $(\frac{1}{3},\frac{1}{3})$
from spin wave excitations of the stripe superlattice. The other
scan in Fig.\,\ref{fig:2}(a) records elastic scattering along the
same line, and contains only the Bragg peak due to the ordered
stripe superlattice. This scan shows that the diffuse scattering
does not have a static component. We have also scanned the
2.5\,meV diffuse inelastic scattering parallel to the $(0, 0, l)$
direction at fixed in-plane wavevector (not shown). The intensity
was found to decrease monotonically away from $l=0$ without any
detectable modulation with $l$.

We next describe two measurements made with neutron polarization
analysis. First, we performed scans along lines equivalent to A
and B at a constant energy of 2.5\,meV, recording the spin-flip
(SF) and non-spin-flip (NSF) scattering with the neutron
polarization ${\bf P}$ parallel to the scattering vector ${\bf
Q}$. The B data are shown in Fig.\,\ref{fig:2}(b). In both scans
the diffuse scattering feature was observed only in the SF
channel. We conclude from this that the diffuse scattering is
magnetic in origin \cite{MRK-PR-1969}.

Second, we analyzed the polarization of the 2.5\,meV diffuse
scattering at two different wavevectors, ${\bf Q}_1 = (0.25, 0.25,
3.5)$ and ${\bf Q}_2 = (0.75, 0.75, 0)$. Neutrons scatter from
spin fluctuation components perpendicular to ${\bf Q}$, so at
${\bf Q}_1$ the signal is mainly from in-plane fluctuations and at
${\bf Q}_2$ the signal comes equally from in-plane and $c$-axis
components. For each ${\bf Q}$ we measured the SF intensity with
${\bf P} \parallel {\bf Q}$ and with ${\bf P} \perp {\bf Q}$ in
two orthogonal directions. Because SF scattering arises from spin
fluctuations perpendicular to ${\bf P}$ these measurements allow
us to determine the spin fluctuation anisotropy
\cite{MRK-PR-1969}. From the ${\bf Q}_1$ data we found that the
spin fluctuations are isotropic within the NiO$_2$ layers to
within the experimental error of 20\%. The ${\bf Q}_2$
measurements revealed that the $c$-axis component of the magnetic
response is a factor $2.3\pm 0.4$ larger than the in-plane
response. In other words, the fluctuating spin components that
give rise to the diffuse scattering are anisotropic, with a strong
tendency to fluctuate towards the crystal $c$ axis. This
conclusion is consistent with the observation of a decrease in
scattering intensity in the out-of-plane direction mentioned
earlier.

What can we infer from the results described so far for energies
of 2--2.5\,meV? The existence of ridges of inelastic magnetic
scattering running along the diagonals of the $(h,k)$ reciprocal
lattice plane implies the existence of dynamic spin fluctuations
with a short correlation length along the ridge direction and a
relatively long correlation length perpendicular to the ridge. The
absence of any out-of-plane modulation in the scattering means the
correlations along the $c$ axis are extremely weak. Hence, the
spin fluctuations are quasi-1D, with the strongest correlations
parallel to the diagonals of the square lattice, i.e. parallel to
the stripe directions.

\begin{figure}
\includegraphics{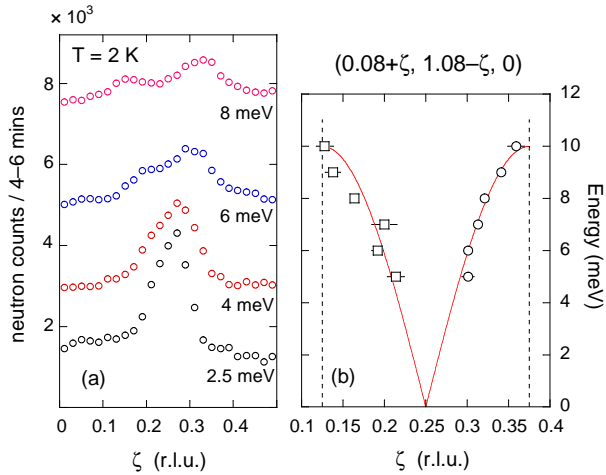}
\caption{(Color online). (a) Constant-energy scans along the line
marked B in Fig.\,\ref{fig:1}(b). Successive scans have been
displaced vertically by 2000 counts for clarity. (b) Dispersion of
the peak positions from a series of constant-energy scans
including those shown in (a). The line is the lower bound to the
excitation spectrum of a spin-$\frac{1}{2}$ Heisenberg AF chain
calculated from Eq.~(\ref{eq1}) with $J=3.2$\,meV. The vertical
dotted lines indicate one period of the lower bound curve.
\label{fig:3} }
\end{figure}

To understand the local pattern of spins responsible for the
diffuse scattering we consider an array of 1D AF chains running
parallel to the direction ${\bf d} = [1, 1, 0]$ with spins
attached to the vertices of the square lattice ---
Fig.\,\ref{fig:1}(c). In the absence of inter-chain correlations
most of the spectral weight is concentrated near the lower bound
of the spin excitation spectrum, which in the case of a
spin-$\frac{1}{2}$ Heisenberg chain is given by
\cite{dCP-PRB-1962}
\begin{equation}
E({\bf Q}) = \pi J|\sin(2\pi{\bf Q}\cdot {\bf d})|, \label{eq1}
\end{equation}
where $J$ is the exchange energy per spin. At vanishingly small
energies, the scattering cross-section is largest when ${\bf
Q}\cdot {\bf d}$ is half an odd integer, and zero when ${\bf
Q}\cdot {\bf d}$ is an integer. Therefore, we expect the
low-energy $(E\ll 4JS)$ scattering to be strongly peaked along the
broken lines shown in Fig.\,\ref{fig:1}(b), which intersect the
scan direction marked A at ${\bf Q} = (\frac{1}{4}, \frac{1}{4},
0)$, $(\frac{3}{4}, \frac{3}{4}, 0)$, etc. Similarly, the
low-energy scattering from spin chains oriented along the other
diagonal ${\bf d}' = [1, -1, 0]$ would lie along the solid
diagonal lines in Fig.\,\ref{fig:1}(b).


It can be seen that the distribution of diffuse scattering we have
observed is consistent this model. The main discrepancy is a
slight meandering of the diffuse scattering shown in
Fig.\,\ref{fig:1}(a). For example, the peak position in the 2\,meV
scan shown in Fig.\,\ref{fig:2}(a) is not quite at
$h=\frac{1}{4}$. Looking in detail we find that the diffuse
scattering closely follows the stripe superlattice zone boundary
\cite{Boothroyd-PRB-2003}, which suggests the spin chains are
weakly coupled and have the same periodicity as the stripes.

We now discuss measurements showing how the diffuse scattering
varies with energy and temperature. In Fig.\,\ref{fig:3}(a) we
plot some of the raw data from a series of wavevector scans along
line B, each scan having a fixed energy. The scan coordinate
$\zeta$ corresponds to $({\bf Q}\cdot {\bf d}')/2$ for a spin
chain parallel to ${\bf d}'$. With increasing energies above $\sim
4$\,meV one can clearly observe two peaks separating from the 1D
AF wavevector $\zeta = \frac{1}{4}$. On Fig.\,\ref{fig:3}(b) we
have plotted the dispersion of these two peaks. The peak centres
were obtained by fitting each of the scans in Fig.\,\ref{fig:3}(a)
with two Gaussians on a linear background. The data are consistent
with a dispersion described by Eq.~(\ref{eq1}) with a maximum
energy of 10\,meV.

\begin{figure}
\includegraphics{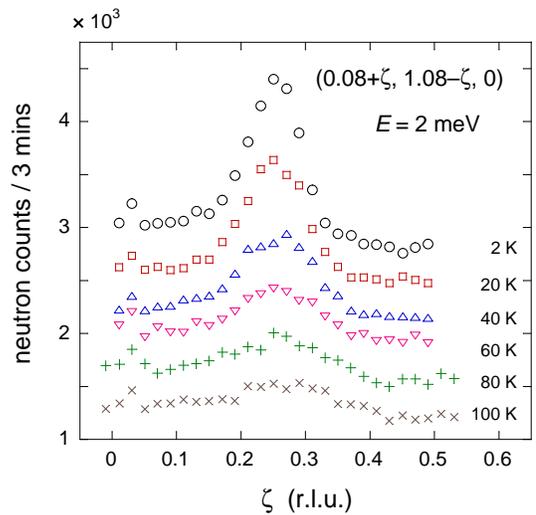}
\caption{(Color online). Temperature dependence of the diffuse
scattering measured in a scan along line B in
Fig.\,\ref{fig:1}(b). The energy was fixed at 2\,meV. Successive
scans have been displaced vertically by 400 counts for clarity.
\label{fig:4} }
\end{figure}

Fig.\,\ref{fig:4} displays the temperature dependence of diffuse
scattering at an energy of 2\,meV, measured once again in a scan
along line B in Fig.\,\ref{fig:1}(b). With increasing temperature
the intensity of the peak decreases and the width increases. These
data show that the spin correlations become weaker with
temperature, and are almost wiped out above 100\,K.

The measurements reported here show that the new spin fluctuations
we have found behave very differently from the low-energy spin
excitations of the host Ni$^{2+}$ spins studied previously
\cite{Boothroyd-PRB-2003}. The latter are localized in $\bf Q$,
have in-plane polarization below an anisotropy gap of $\sim
5$\,meV, and remain correlated up to the magnetic ordering
temperature of $\sim 200$\,K \cite{Lee_Cheong-PRL-1997}. The
diffuse scattering, on the other hand, is spread out in $\bf Q$,
shows predominant out-of-plane anisotropy, and survives only to
$\sim 100$\,K.

These differences imply that the diffuse scattering does not arise
from the AF ordered Ni$^{2+}$ spins located between the charge
stripes, but instead originates from AF correlations in the charge
stripes themselves. Our central conclusion, therefore, is that
there exist quasi-1D AF fluctuations among the stripe electrons.
These fluctuations are dynamic on the timescale probed by neutron
scattering \cite{footnote1}, and from the half-width of the
diffuse scattering ridge we estimate the correlation length of the
fluctuations along the stripes to be $\sim 15$\,\AA.  The charges
are thought to be localized on Ni$^{3+}$ ions \cite{Li-PRB-2003},
which means they would carry a spin $S = \frac{1}{2}$ in the
strong crystal field limit. Hence, we can estimate the AF exchange
parameter $J$ by fitting Eq.~(\ref{eq1}) to the data in
Fig.\,\ref{fig:3}(b). The result is $J\approx 3.2$\,meV.

Attempts to explore the properties of stripe phases theoretically
typically model the stripes as 1D electron liquids coupled via
pair hopping or spin exchange to an insulating AF background
\cite{Stripes-SC-theory}. These models commonly predict a spin gap
in the spectrum of the stripe electrons, allowing various
instabilities to occur, including superconductivity, depending on
the model parameters. The results described here represent the
first measurements of the spin dynamics of stripe electrons, and
are important because they provide data with which to constrain
the above-mentioned charge-stripe models. Hence, this is a step
towards understanding the role of stripes in the problem of
cuprate high-$T_{\rm c}$ superconductivity.

Of course, one should be aware that stripes in the nickelates and
cuprates differ somewhat in character. In the nickelates the
stripes are very stable and are aligned along the diagonals of the
square lattice, whereas in the superconducting cuprates the
stripes are more dynamic and are aligned horizontally and
vertically\cite{Tranquada-Nature-1995}. Also, the 1D band formed
by the stripe electrons is half-filled in the nickelates (hence
insulating stripes) \cite{Zaanen-Littlewood-PRB-1994}, but is
quarter-filled in the cuprates (metallic stripes). Nevertheless,
the discovery of AF correlations among the stripe electrons in a
nickelate (in contrast to theoretical predictions of ferromagnetic
stripes \cite{Zaanen-Littlewood-PRB-1994}) raises general
questions about the mechanism of intra-stripe spin exchange and
about the coupling between the stripe electrons and the
surrounding AF region. A grasp of these phenomena in a simple
reference material such as La$_{5/3}$Sr$_{1/3}$NiO$_4$ could
contribute towards a better understanding of the stripe phase in
other systems. Looking ahead, our results suggest it would be well
worthwhile conducting a search for spin correlations among the
stripe electrons in a stripe-ordered cuprate.

We thank L.-P. Regnault for help with the experiments on IN22.
This work was supported by the Engineering \& Physical Sciences
Research Council of Great Britain, and was performed in part at
the Swiss Spallation Neutron Source SINQ, Paul Scherrer Institute
(PSI), Villigen, Switzerland.

\end{document}